%Paper: cond-mat/9302037
%From: George Batrouni <ggb@Think.COM>
%Date: Thu, 25 Feb 93 07:42:42 EST

% tex file needing the jnl3.tex and reforder.tex macros
%
\def \pm {{+ \over }\,\,}
\input jnl3.tex
\input reforder.tex
% *********************************************
%  macros for the table
%
% shorthand
%
%

%
%
% here is Knuth's box:
%
\def\boxit#1{\vbox{\hrule\hbox{\vrule\kern3pt
      \vbox{\kern3pt#1\kern3pt}\kern3pt\vrule}\hrule}}
%
% here is a kludge box:
%
\def\kluboxit#1{\vbox{\hrule\hbox{\vrule
      \vbox{\kern3pt#1\kern3pt}\vrule}\hrule}}
%
% here are things for text not to go into the table columns
%
%    single horizontal rule
%
     
%
%   double horizontal rule
%
     
%
%
     \def\rullesud{\noalign{\vskip3pt} \noalign{\hrule} \noalign{\vskip3pt}}
     \def\rullesu{\noalign{\vskip3pt} \noalign{\hrule}}
     \def\rullesd{\noalign{\hrule} \noalign{\vskip3pt}}
%
%    any general `aside'
%
     
%
%   here are some common TeX templates
%
%    l <=> left
%    r <=> right
%    c <=> center
%    j <=> justified
%    q <=> quad at end
%    Q <=> quad at beginning
%
%
%
     
%
     
%
     
%
     \def\cj{\hfil ## \hfil}
%
     
%
     
%
     
%
     
%
     
%
% any more?
%
%
%
%  null character to align decimal points
%  always use nulit inside a group so the character goes back to
%  standard usage
%
%   \nulit?
%              ??0.12?
%              137.0??
%              ???.3??
%
                   \newdimen\digitwidth
                   \setbox0=\hbox{\rm0}
                   \digitwidth=\wd0
%
%
%  ok,  for now we just
%  have this everywhere
%  and use mtable#1
%  `?' can't be used in the tables
%
                   \catcode`?=\active
                   \def?{\kern\digitwidth}
    \def\nulit#1{  \catcode`#1=\active
                   \def#1{\kern\digitwidth}   }

    \def\nulqit{  \catcode`?=\active
                   \def?{\kern\digitwidth}   }
%
%
% here are abbrevs for $$\vbox.....
%
   
%
%
   
%
   
%
   
%
%
   
%
%
%  here is shorthand for plus or minus
%
   
%
% end macros for table
% ****************************************************

\title{\bf Universal Conductivity in the
Two dimensional Boson Hubbard Model}

\author G.G. Batrouni,$^{1}$, B. Larson$^{1}$,
R.T. Scalettar,$^{2}$ J. Tobochnik$^{3,4}$, J. Wang$^{4}$

\affil
$^{1}$Thinking Machines Corporation
245 First Street
Cambridge, MA 02142
\affil
$^{2}$Physics Department
University of California
Davis, CA 95616
\affil
$^{3}$Physics Department
Kalamazoo College
Kalamazoo, MI 49006
\affil
$^{4}$Physics Department
McGill University
Montreal, PQ Canada H3A 2T8

\abstract
We use Quantum Monte Carlo to evaluate the conductivity $\sigma$ of
the 2--dimensional
disordered boson Hubbard model at the superfluid-bose glass phase
boundary. At the critical point for particle density $\rho=0.5$,
we find $\sigma_{c}=(0.45 \pm 0.07) \sigma_{Q}$, where $\sigma_{Q}=
e_{*}^{2} / h$ from a finite size scaling analysis of the
superfluid density.  We obtain $\sigma_{c}=(0.47 \pm 0.08) \sigma_{Q}$
 from a direct calculation of the current--current correlation function.
Simulations at the critical points for other particle densities,
$\rho=0.75$ and $1.0$, give similar values for $\sigma$.
We discuss possible origins of the difference in this value
 from that recently obtained by other numerical approaches.

\endtitlepage

\centerline{{\bf Introduction}}
\vskip0.2in

The interplay between interactions and disorder has been a compelling
field of research over the last decade.  Particular interest has
focused on two dimensions, both because randomness alone marginally
localizes the eigenstates, and also because of a set of fascinating
experiments on the superconducting--insulator transition.
Early studies\refto{ORR} of the sheet resistance as a function of temperature
for granular systems of Sn and Ga on glass substrates suggested that
the resistance was the only important variable controlling
whether the system went insulating or superconducting at low temperatures.
Films differing significantly in thickness or other characteristics,
but having similar resistance,
would end up in the same low $T$ ground state phase.
Furthermore, it was found that the transition to the superconducting state
consistently took place at a resistance close to $h / 4e^{2} = 6.45 k\Omega$.
On the other hand,
experiments\refto{HAVILAND} on homogeneously disordered Bi and Pb films
on Ge substrates explicitly employed the thickness as a control
parameter-- for thin films the resistivity is characteristic
of an insulator, i.e. $\rho$ increases as the temperature is lowered,
while for thick films a superconducting phase transition is observed.
In between is a separatrix which has the property
that, roughly speaking, $\rho$ neither
diverges nor goes to zero  as $T \rightarrow 0$.
For Bi the value of the resistance
at the critical thickness is similar to that found for granular systems,
namely close to $6.45 K\Omega$.
For Pb the value is somewhat larger, around $9.5 K\Omega$.
Since these materials were homogeneously
disordered, it was suggested
that the universal resistance phenomenon must be attributed to
a more general principle than those originally suggested to
explain the behavior of strictly granular materials.

Meanwhile the mechanism for the destruction of superconductivity
was also explored.  In granular Al,
Pb, and Sn films, it was found\refto{DYNES1} that the
superconducting gap and $T_{c}$ varied rather little with thickness.
The insulating transition was driven by long resistive tails which
gradually broadened until they reached the size of the gap itself.
For homogeneous Pb films, on the other hand,
$T_{c}$ and the gap were driven to zero by the disorder.\refto{DYNES2}
This work emphasized the possibility of destroying
superconductivity either through a reduction of the amplitude
of the pair wave function or alternately through the
destruction of phase coherence.\refto{DYNES2}

Since then, many experimental groups have explored related
effects for different materials and different control parameters.
Ga was observed\refto{JAEGER} to have a critical resistance of
$\approx 6 K\Omega$.  In $MoC$ films, again using the thickness
to dial through the transition,\refto{LEE} a threshold resistance
in the range $2.8-3.5 K\Omega$ was measured.  In doped, amorphous
semiconductor films (InO$_{x}$) of fixed thickness, the degree of
microscopic disorder has been used to dial the transition.\refto{HEBARD1}
Also in InO$_{x}$ ($R_{c} \approx 5 K\Omega$) a magnetic field
can drive\refto{HEBARD2} the systems across the
superconducting--insulator phase boundary.
Finally, in Josephson junction arrays\refto{GEERLIGS}
the ratio of charging to Josephson energies is similarly tuned.

While the experiments all describe qualitatively similar behavior,
the quantitative situation is still developing.
Not only does the ``universal conductance'' in fact vary,
but there can be non--trivial structure in the
curves near the separatrix as the temperature is lowered.
In particular, a ``re--entrant'' phenomenon is observed
in which the resistivity dips as if the film were about to go superconducting,
but then the transition is usurped by the formation of an insulating
state and the resistance rises as $T$ is decreased further.\refto{ORR}
Further questions concern whether the experiments are
really in the critical regime or not.  That the experimentally
accessible temperature ranges, $ T \ge 0.5 K$, may not
be sufficiently low is an issue that has been raised by,
among other things, the
existence of structure in the curves as $T$ is reduced.
This is a crucial question in determining the appropriate model
for the critical phenomena, as we shall discuss further.

There have been a number of theoretical efforts
to understand these phenomena.\refto{THEORIES}
A particularly interesting set of ideas has centered on the proposal
that, despite the fact that the underlying degrees of freedom
are fermionic, the appropriate model is one of
disordered, interacting bosons,\refto{FISHER1,THEORIES}
$$
H=-{t \over 2}\sum_{\langle ij \rangle}  (a_{i}^{\dagger}a_{j}+
a_{j}^{\dagger}a_{i}) + {V \over 2} \sum_{i} n_{i}^{2}
+\sum_{i} \epsilon_{i} n_{i}
\eqno(1)
$$
Here $a_{i}^{\dagger}$ is a boson creation operator
at site $i$, $n_{i}=a_{i}^{\dagger}a_{i}$ is the number operator,
$V$ is a soft--core on--site interaction, and $\epsilon_{i}$ is
a random local chemical potential, which we chose
to be uniformly distributed between $-\Delta$
and $\Delta$. The physical motivation for a boson
model is provided by
imagining one has pre--formed Cooper pairs {\it above} the superconducting
transition temperature, i.e. the fermions first condense into a set of
interacting bosons that lack phase coherence. At lower temperatures
phase coherence is established, and the system becomes superconducting.
Such a view is particularly natural for granular systems where one can
imagine Cooper pairs forming on individual grains.
Cha {\it et.al.}\refto{CHA} have presented a more careful
characterization of this general scenario by
comparing correlation lengths for the various
operator expectation values associated with the propagation
of two electrons both together and independently.
The fundamental physics is
that on the scale of the diverging pair--correlation length
Cooper pairs appear as point bosons.

However,
the justification for considering the Hamiltonian Eq.~1 is
by no means completely qualitative.
The renormalization group calculation of Giamarchi
and Shultz\refto{GIAMARCHI}
provides at least one explicit {\it theoretical} demonstration
that the universality classes describing the superconducting
transition of fermions with an attractive interaction and the
superfluid transition for disordered bosons are identical.
More detailed {\it experimental} justification of
a picture of pre--formed bosons has been provided by
the work of Paalanen {\it et.al.}\refto{PAALANEN},
which lends rather compelling support to a picture where
the amplitude of the Cooper pair
wave function is finite on both sides of $T_{c}$, and the superconducting
transition is indeed described in terms of a loss of
phase coherence between pairs rather than the
breaking of the pairs themselves.
Even so, it must be pointed out that tunneling
experiments\refto{DYNES2,VALLES} on homogeneously disordered Pb
films have indicated a vanishing of the gap at $T_{c}$.
It therefore may be that granular and Josephson junction systems are
more appropriate realizations of bosonic models.

Here we will describe the results of quantum simulations combined with a
finite size scaling analysis to determine the conductivity at the
superfluid--insulator transition in the boson--Hubbard model.  The
organization of this paper is as follows: We will first review briefly some
of the analytic and numerical work to put our calculation in context.  Next
we describe our monte carlo method and discuss how the conductivity can be
obtained either from a Kubo formula for the current--current correlation
function, or, alternately, from the rigidity of the system to phase
changes.  We then detail our numerical results, beginning with the
determination of the critical point for the bose glass--superfluid
transition, and continuing with the analysis yielding the conductivity.
The two techniques yield values which are the same to within our estimated
error bars.  However, this number differs substantially from that obtained
previously.\refto{RUNGE,SORENSEN} We conclude with a discussion of some
possible sources of this discrepancy.

Fisher {\it et.al.}\refto{FISHER1} have qualitatively mapped out the ground
state phase diagram of the boson-Hubbard model.  In the clean limit, a
gapless superfluid phase exists for all non--integer fillings. At
commensurate densities, however, the interactions freeze the bosons into a
gapped Mott insulating (MI) phase for sufficiently strong coupling.
Increasing hybridization $t$ will eventually wash out the gap and drive
the system from insulator to superfluid.  It was argued that this
``coupling--driven'' transition is in the universality class of the
classical $d+1$ dimensional XY model.  By contrast, changing the density
away from an integer number of bosons per site can also induce
superfluidity, but here the transition is mean field in character.  When
disorder is added, it was suggested\refto{FISHER1} that a third, ``Bose
glass,'' phase appears.  This phase is characterized by the absence of a
gap, but also by a vanishing superfluid density.  While all non--integer
boson densities were superfluid prior to the introduction of randomness, an
incommensurate insulating phase is now possible as the disorder increases.
This work\refto{FISHER1,FISHER3} also predicted values or bounds for the
critical exponents, some of which have been verified
experimentally.\refto{HEBARD2}

The ground state phase diagram was subsequently mapped out
numerically\refto{US1,US2,US3,NANDINI,SINGH} in one and two dimensions and
also studied by the Bethe Ansatz.\refto{KRAUTH} Quantitative values for the
coupling required to localize the bosons into the MI phase, were
determined,\refto{US1,NANDINI,SINGH} and the prediction of mean--field
exponents for the density controlled transition was verified.\refto{US1} In
the presence of disorder, the basic picture of the formation of a bose
glass phase was substantiated, although a variety of unexpected re-entrant
phenomena were also oberved.\refto{US2,NANDINI}

Numerical and analytic efforts have more recently turned to the transport
properties for two dimensional lattices, and, in particular, the evaluation
of the conductivity.  Runge used exact diagonalization techniques, combined
with finite size scaling to extract the conductivity of the disordered,
hard--core ($U=\infty$) model.  He found $\sigma_{c}=(0.15 \pm
0.01)\sigma_{Q}.$ In a set of papers,\refto{CHA,SORENSEN,OTHERS} the 3--d
XY model and its Villain variant were studied with and without disorder and
with short and long range potentials.  It was argued that this model is in
the same universality class as the disordered boson--Hubbard Hamiltonian.
Cha {\it et.al.}\refto{CHA} obtained the conductivity first in the clean
system with short--range interactions.  The value $\sigma_{c}=(0.29 \pm
0.02) \sigma_{Q}$ found by Monte Carlo compared favorably with an analysis
based on a $1/N$ expansion which gave $\sigma_{c}=0.251 \sigma_{Q}$.
Sorensen {\it et.al.}\refto{SORENSEN} then showed that the addition of
randomness results in a {\it smaller} value $\sigma_{c}=(0.14\pm 0.01)
\sigma_{Q}$, that is, farther from the experimentally realized numbers.
Finally,\refto{SORENSEN} the disordered model with a long range Coulomb
potential was studied.  Including these interactions was found to push the
conductivity back up to $\sigma_{c}=(0.55 \pm 0.01) \sigma_{Q}$.  This
final value is certainly within a factor of two or so of the experiments,
and possibly substantially closer especially considering uncertainties
associated with the precise low temperature experimental values.

None of these studies were of the Hamiltonian Eq.~1.
It is of interest to compute
the properties of the boson--Hubbard model directly,
including the effect of the number fluctuations ignored in
the mapping to the XY model and also in the
hard--core diagonalization methods.
One motivation is to test the arguments suggesting the universality classes
are identical.  In addition, if the experiments are not
in the critical regime, then {\it non--universal}
quantities become of interest, and the predictions of the
original model are essential.

\vskip0.2in
\centerline{{\bf Monte Carlo and Finite Size Scaling Methods}}
\vskip0.2in
Here we present a brief discussion of our numerical approach.
More detailed descriptions
have recently appeared.\refto{WORLD,US3}
We begin by expressing the partition function as a path integral.
In order to do this, we discretize the imaginary time
$\beta=L_{\tau}\Delta \tau$ and use the Trotter
approximation\refto{TROTTER}
to decompose the imaginary time evolution operator.
$$
Z={\rm Tr}\,\,e^{-\beta H} = {\rm Tr}\,\,[e^{-\Delta \tau H}]^{L_{\tau}}
\approx {\rm Tr}\,\,[\prod_{i}e^{-\Delta \tau H_{i}}]^{L_{\tau}}.
\eqno(2)
$$
This is a well--controlled procedure since one can explicitly
calculate at different $\Delta \tau$ and use well
understood techniques\refto{FYE} to extrapolate to $\Delta \tau =0$.
We now insert complete sets of states
to express $Z$ as a sum over a {\it classical} occupation number field
$n(\vec l,\tau)$.
$$
Z= \sum_{\{n(\vec l,\tau)\}}
\langle n(\vec l,1 | e^{-\tau H_{1}} | n(\vec l,2 \rangle
\langle n(\vec l,2 | e^{-\tau H_{2}} | n(\vec l,3 \rangle
\,.\,.\,.\,.
\eqno(3)
$$
We have thus written our $d$ dimensional quantum mechanical trace as a
classical statistical mechanics problem in $d+1$ dimensions.
The classical degrees of freedom to be sampled
are the space--imaginary time
dependent boson occupation number field.  Due to particle number
conservation in $H$, the allowed configurations of this field trace
out ``world--lines'' as the particles propagate in $\tau$.
The ``Boltzmann weight'' of a particular configuration
is the product of matrix
elements, and can be sampled with standard stochastic techniques.
Various choices are possible for the decomposition of $H$ into
$H_{i}$.  We chose the ``checkerboard'' break--up in these
studies.\refto{BARMA}

Expectation values are constructed in a similar way.
Of particular interest to us will be the true (paramagnetic)
current--current correlation function ${\cal J}_{xx}^{p}(\tau)$
$$
\eqalign{
{\cal J}_{xx}^{p}(\tau)&=\langle j_{x}^{p}(\tau)j_{x}^{p}(0) \rangle
\cr
j_{x}^{p}(\tau)&=e^{H\tau} j_{x}^{p}(0) e^{-H \tau}
\cr
j_{x}^{p}(0)&=it \sum_{\vec i} (a_{\vec i + \hat x}^{\dagger} a_{\vec i} -
a_{\vec i}^{\dagger} a_{\vec i+\hat x}),
\cr}
\eqno(4)
$$
and its fourier transform
$$
{\cal J}_{xx}^{p}(iw_{m})=\int_{0}^{\beta} d \tau \langle
j_{x}^{p}(\tau) j_{x}^{p}(0) \rangle
e^{i \omega_{m} \tau}.
\eqno(5)
$$

If we were able to evaluate these quantities at real frequencies,
$\omega$, we could then compute the frequency dependent
conductivity from the Kubo formula \refto{MAHAN,SCALAPINO1,SCALAPINO2}
$$
\sigma(\omega)=  \sigma_{Q} {2 \pi \over \omega}
[  -{\cal J}_{xx}^{p}(\omega) - \langle k_{x} \rangle ].
\eqno(6)
$$
Here $k_{x}$ is the kinetic energy in one of the two equivalent lattice
directions.  The dc conductivity is then obtained from $\sigma_{{\rm dc}} =
\sigma(\omega\rightarrow 0)$. However,
it is well--known that the analytic continuation of quantities like
the conductivity to real frequencies is a subtle problem.\refto{MAXENT}
Following Sorensen {\it et.al.},\refto{SORENSEN}
our approach will be to exhibit that for small Matsubara frequencies
($\omega_m = 2m\pi/\beta$), the conductivity obeys
$\sigma(i \omega_{m}) = \sigma_{*} / (1+ |\omega_{m}| \tau_{c})$.
With the empirical observation
that our data fit this analytic form, the
continuation to $\sigma(\omega + i \delta) = \sigma_{*}
/ (1 - i \omega \tau_{c})$ is then possible.
A direct calculation of $\sigma$, avoiding the
assumption of the Drude form, similar to the
computation of real frequency quantities
in fermion simulations,\refto{JARRELL} would be useful.

There is an alternate approach to obtaining the
conductivity based, instead, on the response of the system to
a phase twist.  We define a ``pseudocurrent'' operator and its correlation
function by
$$
\eqalign{
\tilde {\cal J}(\tau) &= {1 \over {\beta L^2} } \sum_{\tau^\prime} \langle
\tilde j_x(\tau+\tau^\prime) \tilde j_x(\tau^\prime) + \tilde
j_y(\tau + \tau^\prime) \tilde j_y(\tau^\prime) \rangle,
\cr
\tilde j_x(\tau)&= \sum_{i=1}^{N_{b}} [x(i,\tau+1) - x(i,\tau)].
\cr}
\eqno(7)
$$ Here $x(i,\tau)$ is the $x$--component of the position of boson $i$ at
time slice $\tau$, $N_b$ is the total number of bosons and $L$ the extent
of the lattice in each spatial direction.  Therefore, $\tilde j_x(\tau)$
measures the number of bosons moving in the positive $x$ direction minus
those moving in the negative direction at time slice $\tau$, i.e. the net
boson flux per time step at time slice $\tau$ in the $x$ direction. Similar
definitions apply for the pseudocurrent in the $y$ direction. This
pseudocurrent operator was originally introduced\refto{US1} to measure the
mean square winding number and from that the superfluid
density.\refto{POLLOCK}
$$
\eqalign{
\tilde {\cal J}(\omega)&=\sum_{\tau}e^{i \omega \tau} \tilde {\cal J}(\tau),
\cr
\tilde {\cal J}(\omega\rightarrow 0)&= { 1 \over \beta}\langle W^{2} \rangle
\cr
\rho_{s} &= {1 \over 2t} \tilde {\cal J}(\omega\rightarrow 0).
\cr}
\eqno(8)
$$
Defining $\rho_s(\omega)=\tilde {\cal J}(\omega)/2t$, we can show that
the frequency dependent conductivity can be written as
$$
{\sigma(\omega) \over \sigma_{Q}} = 4 \pi t {\rho_{s}(\omega) \over \omega}.
\eqno(9)
$$
Again a Drude form must be assumed here to carry out the analytic
continuation.

Clearly, these two approaches are related, since they both yield the
conductivity.  Indeed, a detailed discussion of the connections between the
response of the free energy to changes in the boundary conditions and the
current--current correlation function has recently
appeared.\refto{SCALAPINO2}  However, despite these general relationships,
the formulations of Eqs.~4--6 and ~7--9 allow us to evaluate $\sigma$ in
two quite different ways.  In particular, the current--current correlation
function is a simple operator expectation value for the model described by
the boson--Hubbard Hamiltonian.  Meanwhile, the pseudocurrent analysis is
based on topological properties of the boson world lines in our path
integral representation of the partition function.  Thus we regard the two
measurements as rather independent determinations of the conductivity and a
check for the self consistency of our simulations.

It is useful to contrast our Monte Carlo approach with the two other
numerical techniques already used in evaluating the conductivity.  As
discussed above, Runge\refto{RUNGE} used an exact diagonalization method
for the $U=\infty$ model.  The advantage of that approach is, most
importantly, that real time correlation functions can be directly inferred,
without the need for any analytic continuation.  A second advantage is that
there are no statistical errors from Monte Carlo sampling, only
fluctuations associated with the disorder averaging.  On the other hand,
the technique is limited to rather small lattices ($2 \times 2$ up to $5
\times 5$) since the Hilbert space dimension grows exponentially with the
number of sites.  Indeed, the Hilbert space also grows rapidly with the
number of bosons allowed per site, so that in practice it is necessary to
consider the hard core case where the site occupations are limited to
$0,1$.  While this prevents the study of the transition as a function of
interaction strength, as we will do, one is able to evaluate the
conductivity for $U=\infty$ at a critical point accessed by changing the
density.

Meanwhile, Monte Carlo simulations have also been conducted of a
spin model (the classical 3--d XY model)
argued to be in the same universality class as the boson
Hubbard Hamiltonian.  These studies are very close in spirit to the
ones described here, since they are also simulations
of a classical model in one higher dimension than the
original quantum Hamiltonian.  Unlike diagonalization, they are
subject to error bars associated with statistical sampling
and have to argue the analytic continuation,
but they are able to study lattices of order 100 sites.
There are two differences with our approach.
The action in our simulation, the product of matrix elements of
Eq.~3 is rigorously appropriate to the boson--Hubbard Hamiltonian.
For example, the mapping to the spin--1/2 model
neglects boson number fluctuations.
Presumably, the advantage of the spin model approach is that the
action is somewhat simpler than that arising from the
quantum$\rightarrow$classical ``world--line''
description detailed above.  On the other hand, there are
evidently assumptions
associated with the universality class of the transition and
the relevant degrees of freedom.

\vskip0.2in
\centerline{{\bf Numerical Results}}
\vskip0.2in

In order to evaluate $\sigma$, we must first determine the
superfluid--insulator critical point.
Here we closely
follow the finite size scaling procedure of Ref.~\refto{SORENSEN}.

According to two parameter finite size scaling, physical quantities, such
as $\rho_s$, on different size lattices, of linear extent $L$,
satisfy\refto{SCALE}
$$
\rho_{s} = L^{\alpha} f(a L^{{ 1 \over \nu}} \delta, \beta
L^{-z}).
\eqno(10)
$$
Here $\alpha = 2 -d-z$, $\beta$ is the inverse temperature, and $\delta
= (V-V_{c})/V_c$ measures the distance to the critical point. The function
$f$ is universal but the metric factor $a$ is not. The dynamic critical
exponent $z$ is predicted\refto{FISHER2} to have the value $z=2$, and our
system is two dimensional giving $\alpha=-2$. Notice that there is no
nonuniversal metric factor for the second argument of the function $f$.
Therefore, by keeping the second argument, $\beta L^{-z}$, fixed and
plotting $L^2\rho_s$ versus $V$ for different lattice sizes, all the curves
should intersect at the critical value $V_c$.  We chose the inverse
temperature $\beta = (1 /4) L^{z} \Delta \tau$ for different lattices $L$,
i.e. the same aspect ratio as in Ref.\refto{SORENSEN}.  Data for
$L^{2}\rho_{s}$, obtained from the pseudocurrent correlation function via
the procedure described in Eqs.~7--9, is shown in Fig.~1.  We see that the
curves from different lattice sizes converge at a value $V/t\approx 7.0$,
but then do not fan out. The reason for our inability to resolve the exact
point at which these curves intersect is the uncertainty introduced by the
extrapolation of the superfluid density, $\rho_s$ to $\omega = 0$.  Because
of this, we use Fig.~1 only to give us a rough idea of where the critical
point is. Note that at the critical point, plots of $\rho(\omega)/\omega$
versus $\omega$ for different lattice sizes should collapse on a single
curve (see Eq.~9). We use this collapse of the data onto a single curve to
determine the critical point more accurately than possible with figure 1.

In our simulations at half filling, $\rho = 0.5$, the strength of the
disorder was kept constant at $\Delta = 6$, and the coupling, $V$, varied
to find the critical point. The number of disorder realizations we for most
of our simulations is of the order of 100 (see the figures for more
details). For the $8\times 8$ and $10\times 10$ lattices we did about 40000
thermalization and 100000 measurement sweeps, for the $12\times 12$ lattice
we did 100000 thermalization and 20000 measurement sweeps. We found that
the large number of sweeps was necessary for thermalization and good
statistics.

Figures 2a-c show plots of $\rho(\omega)/\omega$ versus $\omega$ for
different values of the coupling, $V/t = 6.5, 7.0, 8.0$. We see that the
best data collapse is at $V/t=7.0$. Furthermore, the plots show that the
values $6.5$ and $8.0$ actually {\it bracket} the critical region. This is
because for $V/t=6.5$ the data for $8\times 8\times 16$ lie slightly below
those for $10\times 10 \times 25$, as one would expect if the system is in
the superconducting phase, while the pattern is reversed for $V/t=8.0$, as
should happen when the system goes into the insulating phase.  To obtain
the conductivity, we fit the Monte Carlo data using $\sigma(\omega) = a /
(1 + b \omega_{m})$ (solid lines in the figures).  If we use the data for
$V/t = 6.5$ we find $\sigma \approx 0.5\sigma_Q$, whereas for $V/t=7.4$
(not shown) we get $\sigma = 0.33\sigma_Q$. The best data collapse, for
$V/t=7.0$ gives $\sigma = 0.4\sigma_Q$. To check the effect of the finite
time step errors, which are known to be $O(\Delta\tau^2)$, we redid the
simulation for $V/t=7.0$, with the same $\beta$ but twice as many time
slices, i.e. half the $\Delta\tau$ as before. The results are shown in
Fig.~3 and give $\sigma = 0.44\sigma_Q$. Extrapolating to $\Delta\tau
\rightarrow 0$ gives $\sigma/\sigma_Q = 0.47 \pm 0.08$.

We now compute the conductivity from the true current-current correlation
function using the Kubo relation by taking the limit $\sigma(\omega
\rightarrow 0)$ in Eq.~9.  In Fig.~4 we show the fourier transform of the
current-current correlation function, $\tilde{\cal J}(\omega)$, as a
function of $\omega$ at the critical point $V=7.0$ for $\Delta=6$.  for
lattice sizes $N=8 \times 8, 10 \times 10,12 \times 12$.  To obtain the
conductivity, we fit the data with the Drude form
$$
\tilde{\cal J}(\omega)-\tilde{\cal J}(\omega=0)= { A\omega \over 1 +
\omega T }.\eqno(11)
$$
According to the Kubo formula, Eq.~6, the kinetic energy should be
subtracted from the frequency dependent current--current correlation
function to obtain the conductivity.  Scalapino {\it
et.al.}\refto{SCALAPINO1,SCALAPINO2} demonstrated recently that whether or
not the $q=0$ current--current correlations extrapolate to the kinetic
energy $k$ as $\omega \rightarrow 0$ can be used as a signature of the
superfluid--insulator transition in the fermion Hubbard model.  We have
verified that this is true for bosons too because our independent
measurements of the current and the kinetic energy do correctly signal this
transition in agreement with other order parameters such as the superfluid
density and gap.  Indeed, this analysis provides a separate check on our
code.  Fig.~5 is the same as Fig.~4 but with twice as many time slices.

Table 1 shows the values for the fitting parameters $A$ and $T$, as well as
the associated conductivity $2 \pi A= 0.38$--$0.46$ with the range in
values representing changes due to different spatial and imaginary time
lattice geometries.  These numbers are in agreement to within our
systematic and statistical uncertainties with those found from the
independent analysis of the pseudocurrent in the previous section.

To check universality, we did similar simulations for $\rho=0.75$ and
$\rho=1$.  The conductivity for $\rho=0.75$ near the transition is shown in
Fig.~6.  These results were obtained using simulations of $5-20$ thousand
sweeps for thermalization and $20-40$ thousand for averaging. Fifty
realizations of the $8 \times 8$ lattice and fifteen realizations of the
$10 \times 10$ lattice were used. In these simulations $t\Delta/V$ was held
fixed at $3/7$ while varying $\Delta$ and $V/t$.  The transition was
determined by searching for where the conductivity changes from increasing
with size to decreasing with size, i.e. as in Fig. 2a-c. In the Bose glass
phase near the transition we find that for the small lattices we are using
it is difficult to distinguish between data which collapse on the same
curve and data where the conductivity decreases with size. This may be one
of the reasons why previous studies have found lower values for $\sigma$.
Fits of the data in Fig.~6 to the form $\sigma(\omega) = a /(1 + b
\omega_{m})$ show that $\sigma/\sigma_q$ varies from $0.36$ to $0.38$.
The range of values depends on the number of data points used in the fit.
Using all the data gives the lower values; using only the two points
closest to $\omega_m = 0$ (which have larger statistical error bars) gives
the higher values.  At $\rho = 1$ we obtain similar results as shown in
Fig.~7. This is an important fact since it is possible that the Bose glass
phase would not completely cover the Mott lobe, and thus at integer filling
one might find a phase transition from the superconductor directly into the
Mott insulator.  The data were obtained with $10,000$ sweeps for
thermalization and $40,000$ sweeps for averaging. From $60$ to $170$
realizations were used for lattice sizes of $L=8$ and $L=10$ and fixed $V/t
= 10$.  These results show reasonable scaling.  Again, estimates for
$\sigma/\sigma_q$ are in the range $0.3$ to $0.4$. The conductivity results
for $\rho=0.75, 1$ are consistent with those for $\rho=0.5$.

Our most accurate data are for half filling, and these were done on the
Connection Machine CM5 with 32 and 64 processing nodes.  The simulations
for $\rho=0.75, 1$ were done on Silicon Graphics workstations using a
program written independently from the one used for half--filling.

\vskip0.2in
\centerline{{\bf Conclusions}}
\vskip0.2in

In this paper we have measured the conductivity of the disordered
boson--Hubbard model.  We obtain a value $\sigma= 0.46 \pm 0.08 \sigma_{Q}$
which differs from those previously reported for studies of a classical
spin model\refto{SORENSEN} and diagonalization of hard--core boson systems
on small lattices.\refto{RUNGE}
One possible source of this discrepancy is that we noticed in our data that
the scaling curves which are extrapolated to zero frequency to obtain the
conductivity, are still curving upwards towards large values for the larger
lattices. It is therefore possible that for larger lattices than examined
here ($12 \times 12$) the conductivity could be even larger.

The precise source of this discrepancy is still under investigation.
We believe that by evaluating the conductivity by two rather different
approaches we have reduced the possibility of trivial questions of
normalization.  This still leaves the possibility of various numerical
uncertainties, including questions concerning
whether we average over a sufficient number of disorder
realizations, the nature of Trotter errors, etc.  We have
explicitly checked this latter point by running at two values of $\Delta
\tau$ and comparing the results.  We found that this source of systematic
error is small.  We also carefully checked the equilibration of our
lattice.
%Indeed, in a general sense these issues are addressed by the
%abiility to collapse the data on single scaling curves.
%*However, Young mentions checking for metastability by
%replica techniques-  we did not do this.*
The biggest source of uncertainty in our calculation
is clearly in our determination of the critical point.
 From Fig.~2, it seems clear that our value for $\sigma$ is
rather sensitive to this choice, and a range of values
is possible.  However, we do not believe that the previously
reported numbers are consistent with our data.  This would
require a choice of $V_{c}$ clearly inconsistent with any sort
of scaling plot.
We are in the process of carrying out some exact diagonalization
studies to pin down the source of the disagreement with QMC
calculations.

\head{Acknowledgements}
\vskip0.2cm

We would like to acknowledge useful conversations with
A. Peter Young and Gergely Zimanyi.  RTS's work was
supported by National Science Foundation grant
NSF DMR--92--06023 and by the donors of The Petroleum Research Fund,
administered by the American Chemical Society. JT thanks the Physics Department
at McGill University for its hospitality while on sabbatical and also
acknowledges support from the donors of The Petroleum Research Fund,
administered by the American Chemical Society.

\vfill\eject

\head{Figure Captions}

\item{1.}
The scaling variable ($L^{2}\rho_{s}$) vs. the interaction strength, $V$.
The superfluid density, $\rho_s$ is obtained from the pseudocurrent
correlation function via the procedure described in Eqs.~7--8. The
convergence region of the different curves gives the approximate critical
value for $V$.

\item{2.}
Plots of $\rho_{s}(\omega)/\omega$ versus $\omega$ for different values of the
coupling, $V/t = 6.5, 7.0, 8.0$. The best data collapse indicates the
critical point, and that happens for $V/t=7.0$.

\item{3.}
Plot of $\rho_{s}(\omega)/\omega$ versus $\omega$ for coupling $V/t = 7.0$.
Here the number of time steps has been doubled to check for Trotter errors.

\item{4.}
${\cal J}(\omega)$ as a function of $\omega$ at the critical point $V=3.5$
for $\Delta=6$ and lattice sizes $N=8 \times 8, 10 \times 10,12 \times 12$.
The good data collapse again indicates that the system is at the critical
point.

\item{5.}
${\cal J}(\omega)$ as a function of $\omega$ at the critical point $V=3.5$
for $\Delta=6$,  for lattice sizes $N=8 \times 8, 10 \times 10,12 \times
12$.  Here the number of time steps has been doubled (at fixed $\beta$) to
check for Trotter errors.

\item{6.} Scaling plot as in figure 2 but at three quarter filling.
$\sigma/\sigma_q\approx 0.37$.

\item{7.}  Same as figure 6 but at $\rho=1$. $\sigma/\sigma_q\approx 0.35$.

\references

\refis{POLLOCK} E.L. Pollock and
D.M. Ceperley, Phys. Rev. {\bf B30}, 2555 (1984);
D.M. Ceperley and E.L. Pollock, Phys. Rev. Lett. {\bf 56}, 351 (1986); and
E.L. Pollock and D.M. Ceperley, Phys. Rev. {\bf B36}, 8343 (1987).

\refis{MAXENT}  J.E. Gubernatis, M. Jarrell, R.N. Silver,
and D.S. Sivia, Phys. Rev. {\bf B44}, 6011 (1991)

\refis{JARRELL} R.N. Silver, J.E. Gubernatis, D.S. Sivia,
and M. Jarrell, Phys. rev. Lett. {\bf 65}, 496 (1990).

\refis{SCALE}  V. Privman and M. E. Fisher, Phys. Rev. {\bf B30}, 322
(1984).

\refis{OTHERS}  S.M. Girvin, M. Wallin, E.S. Sorensen, and
A.P. Young, Physica Scripta, {\bf VT42}, 96 (1992);
S.M. Girvin, M. Wallin, M. Cha, M.P.A. Fisher, and
A.P. Young, Prog. Theor. Phys.
Suppl., {\bf N107}, 135 (1992).

\refis{WORLD} J.E. Hirsch, R.L. Sugar, D.J. Scalapino and R. Blankenbecler,
Phys. Rev. {\bf B26}, 5033 (1982).

\refis{BARMA}
M. Barma and B.S. Shastry, Phys. Rev. {\bf B18}, 3351 (1978). The two
dimensional version which we used in this study is dicussed in M. S. Makivic
and H.-Q. Ding, Phys. Rev. {\bf B43} 3562 (1991).

\refis{TROTTER}
H.F. Trotter, Proc. Am. Math. Soc. {\bf 10}, 545 (1959);
M. Suzuki, Phys. Lett. {\bf 113A}, 299 (1985).

\refis{FYE}  R.M. Fye,
Phys. Rev. {\bf B33}, 6271 (1986); R.M. Fye and R.T. Scalettar,
Phys. Rev. {\bf B36}, 3833 (1987).

\refis{THEORIES}  For a brief review, see
M. Cha, M.P.A. Fisher, S.M. Girvin, M. Wallin, and
A.P. Young, Phys. Rev. {\bf B44}, 6883 (1991),
and the references cited therein.

\refis{ORR}
B.G. Orr, H.M. Jaeger, A.M. Goldman and C.G. Kuper,
Phys. Rev. Lett. {\bf 56}, 378 (1986).

\refis{JAEGER}
H.M. Jaeger, D.N. Haviland, A.M. Goldman, and B.G. Orr,
Phys. Rev. {\bf B34}, 4920 (1986).

\refis{HAVILAND}  D.B. Haviland, Y. Liu, and A.M. Goldman, Phys. Rev. Lett.
{\bf 62}, 2180 (1989).

\refis{DYNES1}  R.C. Dynes, J.P. Garno,
G.B. Hertel, and T.P. Orlando, Phys. Rev. Lett.
{\bf 53}, 2437 (1984);  A.E. White, R.C. Dynes, and J.P. Garno,
Phys. Rev. {\bf B33}, 3549 (1986).

\refis{VALLES} J.M. Valles, R.C. Dynes, J.P. Garno,
Phys. Rev. {\bf B40}, 6680 (1989).

\refis{LEE}  S.J. Lee and J.B. Ketterson, Phys. Rev. Lett. {\bf 64},
3078 (1990).

\refis{GEERLIGS}  L.G. Geerligs, M. Peters, L.E.M. de Groot, A. Verbruggen,
and J.E. Mooij, Phys. Rev. Lett. {\bf 63}, 326 (1989).

\refis{HEBARD1}  A.F. Hebard and M.A. Paalanen, Phys. Rev. {\bf B30},
4063 (1984);  Phys. Rev. Lett. {\bf 54}, 2155 (1985).

\refis{HEBARD2}  A.F. Hebard and M.A. Paalanen, Phys. Rev. Lett.
{\bf 65}, 927 (1990).

\refis{CHA}  M. Cha, M.P.A. Fisher, S.M. Girvin, M. Wallin, and
A.P. Young, Phys. Rev. {\bf B44}, 6883 (1991).

\refis{PAALANEN} M.A. Paalanen, A.F. Hebard, and R.R. Ruel,
Phys. Rev. Lett. {\bf 69}, 1604 (1992).

\refis{DYNES2}  R.C. Dynes, A.E. White, J.M. Graybeal, and J.P. Garno,
Phys. Rev. Lett. {\bf 57}, 2195 (1986).

\refis{FISHER1}  M.P.A. Fisher, P.B. Weichman, G. Grinstein, and
D.S. Fisher, Phys. Rev. {\bf B40}, 546 (1989).

\refis{FISHER2}  M.P.A. Fisher, G. Grinstein, and
S.M. Girvin, Phys. Rev. Lett. {\bf 64}, 587 (1990).

\refis{FISHER3}  M.P.A. Fisher, Phys. Rev. Lett. {\bf 65}, 923 (1990).

\refis{SORENSEN}  E.S. Sorensen, M. Wallin, S.M. Girvin, and A.P. Young,
Phys. Rev. Lett. {\bf 69}, 828 (1992).

\refis{US1}
G.G. Batrouni, R.T. Scalettar and G.T. Zimanyi, Phys. Rev. Lett.
{\bf 65}, 1765 (1990).

\refis{US2}
 R.T. Scalettar, G.G. Batrouni, and G.T. Zimanyi, Phys. Rev. Lett.
{\bf 66}, 3144 (1991).

\refis{US3}
G.G. Batrouni and R.T. Scalettar, Phys.Rev. {\bf B46}, 9051 (1992).

\refis{NANDINI}
W. Krauth and N. Trivedi,  Europhys. Lett. {\bf 14}, 627 (1991);
N. Trivedi, D.M. Ceperley, and W. Krauth,
Phys. Rev. Lett. {\bf 67}, 2307 (1991).

\refis{SINGH}
K.G. Singh and D.S. Rokhsar, Phys. Rev. {\bf B46}, 3002 (1992).

\refis{KRAUTH}  W. Krauth, Phys. Rev. {\bf B44}, 9772 (1991).

\refis{RUNGE}  K.J. Runge, Phys. Rev. {\bf B45}, 13136 (1992).

\refis{GIAMARCHI}  T. Giamarchi and H.J. Schulz, Phys. Rev. {\bf B37},
325 (1988).

\refis{SCALAPINO1}
D.J. Scalapino, S.R. White, and S.C. Zhang,
Phys. Rev. Lett. {\bf 68}, 2830 (1992), and the reference
which follows, contain a detailed discussion of the issues involved
in measuring the conductivity and superfluid
density in numerical simulations.

\refis{SCALAPINO2}
D.J. Scalapino, S.R. White, and S.C. Zhang, unpublished.

\refis{MAHAN}  G.D. Mahan, {\it Many--Particle Physics},
Plenum, New York, 1981.

\endreferences
\vfill\eject

\centerline{{\bf Table I.}}
$$
\vbox{ \kluboxit{ \offinterlineskip \openup 6pt
       \halign  {
%
% put THE TEMPLATE here:
\span\cj&\span\cj&\span\cj&\span\cj&\span\cj \cr
% DONT REMOVE THE FOLLOWING '%'. If you dont like them remove the
% whole line. NO BLANK LINES (IE. PARAGRAPHS).
   \rullesd
 \qquad $N$ \qquad  & \qquad $\Delta \tau$ \qquad & \qquad $A$
      \qquad      & \qquad $T$ \qquad & \qquad $\sigma$ \qquad \cr
   \rullesud
\qquad $ 8 \times  8 \qquad $  &\qquad  0.222  \qquad & \qquad 0.061  \qquad
& \qquad 0.33  \qquad & \qquad 0.38  \qquad  \cr
\qquad $10 \times 10 \qquad $  &\qquad  0.222  \qquad & \qquad 0.065  \qquad
& \qquad 0.38  \qquad & \qquad 0.41 \qquad   \cr
\qquad $12 \times 12 \qquad $  &\qquad  0.222  \qquad & \qquad 0.068  \qquad
& \qquad 0.40  \qquad & \qquad 0.43 \qquad   \cr
\qquad $ 8 \times  8 \qquad $  &\qquad  0.111  \qquad & \qquad 0.067  \qquad
& \qquad 0.36  \qquad & \qquad 0.42  \qquad  \cr
\qquad $10 \times 10 \qquad $  &\qquad  0.111  \qquad & \qquad 0.072  \qquad
& \qquad 0.42  \qquad & \qquad 0.45 \qquad   \cr
   \rullesu
%
% up there
%
                                                           }   }   }
$$

\noindent{\underbar{Table 1 Caption:}}
The values of the fitting parameters $A$ and $T$ used to obtain
the smooth curves in Figs.~4a--c, and the associated number
for the conductivity.  $A$ and $T$ were obtained by fitting the
analytic form Eq.~11 to the Monte Carlo data at $\omega=1$ and $\omega=2$,
a process which results in an excellent fit even out to much higher
frequencies.  The imaginary time dimension has been
chosen as $L=N^{2}/4$ for $\Delta \tau=2/9$ and
chosen as $L=N^{2}/2$ for $\Delta \tau=1/9$,
so that the inverse temperature
$\beta$ is the same for the two different discretization intervals.
We have also done least squares fits to the
entire frequency range, a process which generally
gave values for the conductivity which were higher by 10\% or so.
However, we believe that fitting to the low frequency data is the
more appropriate procedure, since the analytic form is expected
to be most valid there.
\vfill\eject

\end